\documentclass[]{elsarticle}
\RequirePackage{amsthm,amsmath,amssymb,bbm}

\usepackage{orcidlink}

\usepackage{rotating}
\usepackage{colortbl}
\usepackage{multirow}
\usepackage{makecell}

\theoremstyle{plain}
\newtheorem{theorem}{Theorem}
\newtheorem{assume}{Assumption}
\newtheorem{lemma}{Lemma}

\newtheorem{prop}{Proposition}[section]

\allowdisplaybreaks

\begin{document}

\begin{frontmatter}
\title{Handling bounded response in high dimensions: a Horseshoe prior Bayesian Beta regression approach}		
\author[aaatienmt]{The Tien Mai\,\orcidlink{0000-0002-3514-9636} }\ead{the.tien.mai@fhi.no}
		
\affiliation[aaatienmt]{
organization={
Norwegian Institute of Public Health}, 
city={Oslo},postcode={0456}, 
country={Norway}}

\begin{abstract}
Bounded continuous responses—such as proportions—arise frequently in diverse scientific fields including climatology, biostatistics, and finance. Beta regression is a widely adopted framework for modeling such data, due to the flexibility of the Beta distribution over the unit interval. While Bayesian extensions of Beta regression have shown promise, existing methods are limited to low-dimensional settings and lack theoretical guarantees. 
In this work, we propose a novel Bayesian approach for high-dimensional sparse Beta regression framework that employs a tempered posterior. Our method incorporates the Horseshoe prior for effective shrinkage and variable selection. Most notable, we propose a novel Gibbs sampling algorithm using Pólya–Gamma augmentation for efficient inference in Beta regression model. We also provide the first theoretical results establishing posterior consistency and convergence rates for Bayesian Beta regression. Through extensive simulation studies in both low- and high-dimensional scenarios, we demonstrate that our approach outperforms existing alternatives, offering improved estimation accuracy and model interpretability.

Our method is implemented in the R package ``betaregbayes" available on Github.
\end{abstract}
		
\begin{keyword}
Beta regression, 
sparsity, 
bounded response, 
posterior concentration rates, 
Gibbs sampler, 
Horseshoe prior.		
\end{keyword}
		
\end{frontmatter}

\section{Introduction}

Continuous bounded or proportion type responses---values bounded between 0 and 1---are commonly encountered across a wide range of scientific fields, for example: meteorology and climatology \cite{mullen2013beta},  chemometrics \cite{karlsson2020liu}, insurance \cite{gomez2014log}, medical research \cite{meaney2014monte}, and education \cite{cepeda2013spatial}. 
Typical examples include vegetation cover fraction, mortality rates, body fat percentage, the proportion of family income allocated to health plans, loss given default in finance, and efficiency scores derived from data envelopment analysis. 
A frequent objective in these contexts is to model the relationship between such bounded/proportion response variables and a set of covariates, either for prediction or interpretation.
Traditional linear regression models are generally unsuitable for this task, as they do not respect the inherent constraints of the data and may yield predictions outside the unit interval. Models based on normal error structures can result in biased estimates and implausible predicted values. Consequently, considerable effort has been devoted to developing regression frameworks that account for the bounded nature of proportion data.

Among the available approaches, the Beta regression model introduced in \cite{paolino2001maximum,kieschnick2003regression,ferrari2004beta} has emerged as the most widely adopted due to the flexibility
of the Beta distribution in modeling data confined to the (0,1) interval. This model and its extensions offer a natural way to model mean and dispersion structures in proportion data. Alternative distributions and modeling frameworks have also been proposed to handle specific characteristics of bounded data, including the simplex \cite{barndorff1991some, Zhang2014regrsarticle}, rectangular beta \cite{bayes2012new}, log-Lindley \cite{gomez2014log}, CDF-quantile \cite{smithson2017cdf}, and generalized Johnson distributions \cite{lemonte2016new}.

Bayesian methods for beta regression have received increasing attention in recent years. Notable contributions include the mixed-effects beta regression model of \citet{figueroa2013mixed}, Bayesian beta regression with unknown support proposed by \citet{zhou2022bayesian}, and robust Bayesian formulations by \citet{figueroa2022robust}. A comprehensive review contrasting Bayesian and frequentist approaches to beta regression is provided in \citet{liu2018review}. However, theoretical developments in this area remain limited—there is currently no formal theoretical treatment of Bayesian beta regression, such as consistency or convergence guarantees.

Importantly, existing Bayesian beta regression models primarily address low-dimensional settings, where the number of predictors is modest relative to the sample size. In the era of big data, such constraints are increasingly unrealistic. With the rise of high-dimensional datasets—where the number of covariates can be large or even exceed the sample size—there is a clear need to adapt beta regression methodologies to accommodate sparsity and regularization. These adaptations must also preserve the unique challenges of modeling bounded responses.
To the best of our knowledge, the recent work of \citet{liu2024beta} is the first to extend beta regression to the high-dimensional case using a dimension reduction technique. However, that approach does not incorporate sparsity, which limits its interpretability and efficiency in very high-dimensional applications.

To bridge this gap, we propose a novel high-dimensional sparse beta regression framework based on a generalized Bayesian approach—an extension of standard Bayesian inference that tempers the likelihood using a fractional power \citep{bhattacharya2016bayesian}. 
This methodology, which has not previously been explored in the context of beta regression, is particularly well suited to high-dimensional inference.
From a theoretical standpoint, the fractional posterior enables refined concentration results \citep{alquier2020concentration}. Our work draws on and extends recent advances in generalized Bayesian learning \citep{bissiri2013general}, and connects with current trends in scalable inference, including variational approximations for fractional posteriors \citep{Knoblauch} and empirical Bayes-inspired calibration methods \citep{martin2020empirical}. For recent advances and applications of fractional posteriors, the reader may refer to \cite{mai2024concentration,mai2025concentration,mai2025optimal}.

Despite the expanding literature, theoretical guarantees for Bayesian beta regression remain conspicuously absent. While significant progress has been made in understanding posterior behavior in high-dimensional linear models \citep{castillo2015bayesian} and generalized linear models (GLMs) \citep{jeong2021posterior}, these frameworks do not directly extend to beta regression due to its non-membership in the natural exponential family. To address this foundational limitation, we establish both posterior consistency and convergence rates for our proposed method. These results represent, to the best of our knowledge, the first theoretical guarantees for Bayesian beta regression, and contribute to a deeper understanding of Bayesian inference for bounded outcome models.

From a computational perspective, our work introduces several important innovations. Existing Bayesian beta regression models typically rely on Metropolis–Hastings within MCMC for posterior inference. In contrast, we develop the first Gibbs sampling algorithm for beta regression by leveraging the Pólya–Gamma augmentation strategy \citep{polson2013bayesian}, which simplifies posterior sampling by enabling full conditional updates for the regression coefficients.

To handle sparsity in high-dimensional settings, we incorporate the Horseshoe prior \citep{carvalho2010horseshoe}, a continuous shrinkage prior known for its favorable properties in sparse regression problems. 
Our work represents the first formal integration of the Horseshoe prior within the beta regression framework. 
We employ the hierarchical scale mixture representation of the Horseshoe prior \citep{makalic2015simple}, yielding a fully Bayesian formulation that captures uncertainty and facilitates shrinkage. This allows our model to selectively identify relevant predictors while controlling for noise, making it especially effective in complex, high-dimensional settings \citep{bhadra2015horseshoe_+, piironen2017sparsity,mai2025hightobit}. 

To evaluate the effectiveness of our proposed Horseshoe method, we carry out extensive numerical studies under both low- and high-dimensional settings. Across all cases considered, our method consistently demonstrates superior performance in terms of  estimation, variable selection and prediction accuracy.
In particular, it outperforms several established state-of-the-art approaches, including maximum likelihood Beta regression and transformed Lasso, highlighting its practical advantages and robustness to model complexity and dimensionality. These findings provide strong empirical support for the theoretical properties of the proposed method.

The remainder of the paper is structured as follows. In Section \ref{sc_model_method}, we introduce Beta regression in high-dimensional settings and outline our proposed Bayesian methodology. Section \ref{sc_gibbs_sampler} presents a novel Gibbs sampling scheme based on the Pólya–Gamma data augmentation framework, tailored to our model. In Section \ref{sc_theory}, we investigate the frequentist properties of the posterior, establishing theoretical guarantees for our approach. Section \ref{sc_numerical} showcases empirical results through simulation studies and a real-world data application. Finally, all technical details and proofs are provided in \ref{sc_proofs}. 

Our method is implemented in the \texttt{R} package ``\texttt{betaregbayes}" available on Github: \url{https://github.com/tienmt/betaregbayes}.

\section{Problem and Method}
\label{sc_model_method}
\subsection{High-dimensional Beta regression}

We study the problem of modeling a continuous outcome variable $y \in (0, 1)$ using a Bayesian Beta regression approach. In this setup, the response is assumed to follow a Beta distribution of the form:
\begin{equation}
    \label{eq_beta_distribut}
y \sim \text{Beta}(\mu \phi, (1 - \mu) \phi),
\end{equation}
where $ \phi $ is the precision and the mean parameter $\mu$ is linked to the linear predictor $\eta = X^\top \beta$ via the inverse logit function, 
$$
\mu 
= 
\text{logit}^{-1}(\eta) = (1 + \exp(-\eta))^{-1}.
$$
Here, $X \in \mathbb{R}^p$ denotes the covariate vector, and $\beta \in \mathbb{R}^p$ represents the regression coefficients. 
Under this model, the expected value of the response is $\mathbb{E}(y) = \mu$. 
The known precision parameter $\phi > 0$, which is common to all observations, controls the variability of the Beta distribution. Our focus is on the high-dimensional regime where the number of predictors $p$ exceeds the number of observations $n$.

In this high-dimensional setting, we assume that the true regression coefficient vector $\beta_0$ is sparse; that is, the number of nonzero entries $s^* := \|\beta_0\|_0$ is smaller than the sample size $n$, which in turn is smaller than the number of covariates $p$. This reflects the assumption that only a limited number of covariates influence the response.

Let $P_\beta$ represent the joint probability distribution of the data pair $(y, X)$ induced by the model parameterized by $\beta$. Let $y_1, \dots, y_n \in (0,1)$ be independent observations, and $X_i \in \mathbb{R}^p $ be the covariate vector for observation $i, i = 1,\ldots,n$. The likelihood over all observations is as:
$$
L_n (\beta ) 
=
\prod_{i=1}^n \left[
\frac{\Gamma(\phi)}{\Gamma(\mu_i \phi) \Gamma((1 - \mu_i)\phi)} \,
y_i^{\mu_i \phi - 1} (1 - y_i)^{(1 - \mu_i) \phi - 1}
\right],
$$
$$
\quad \text{where } \mu_i = \text{logit}^{-1}(X_i^\top \beta).
$$

\subsection{A Horseshoe prior Bayesian approach}

We propose a sparse generalized Bayesian framework for high-dimensional Beta regression. 
Specifically, for a fractional power $\alpha \in (0,1)$ and a sparsity-inducing prior $\pi(\beta) $ defined in \eqref{eq:HS}, we consider the following fractional posterior distribution for $\beta$:
\begin{align}
\label{eq_frac_posterior}
\pi_{n, \alpha}(\beta)
\propto
L_n(\beta)^{\alpha} \pi(\beta),
\end{align}
adopting the notation of \citet{bhattacharya2016bayesian}. When $\alpha = 1$, this reduces to the standard Bayesian posterior. Choosing $\alpha < 1$ introduces a tempered likelihood, which can offer robustness to model misspecification \citep{alquier2020concentration}. In practice, selecting $\alpha$ close to one (e.g., $\alpha = 0.99$) retains the benefits of standard Bayesian inference while improving theoretical properties \citep{martin2017empirical}.

To promote sparsity in the regression coefficients, we adopt the Horseshoe prior \citep{carvalho2010horseshoe}, a well-known global-local shrinkage prior that has gained significant attention in the high-dimensional Bayesian literature for its desirable theoretical and empirical properties. Specifically, we place an independent Horseshoe prior on each regression coefficient $\beta_j$, as follows:
\begin{equation}
\label{eq:HS}
\begin{aligned}
\beta_j \mid \lambda_j, \tau &\sim \mathcal{N}(0, \lambda_j^2 \tau^2), 
\\
\lambda_j &\sim \mathrm{Cau}_+(0, 1), 
\\
\tau &\sim \mathrm{Cau}_+(0, 1),
\end{aligned}
\end{equation}
for $j = 1, \dots, p$, where $\mathrm{Cau}_+(0, 1)$ denotes the standard half-Cauchy distribution, 
truncated to the positive real line, with density proportional to $(1 + u^2)^{-1} \mathbbm{1}_{(0, \infty)}(u)$. We denote the prior induced by this hierarchical specification as $\pi_{HS}$.

The Horseshoe prior is particularly well-suited for sparse high-dimensional problems due to its global-local shrinkage structure. The global parameter $\tau$ controls the overall level of shrinkage across all coefficients, while the local parameters $\lambda_j$ allow for coefficient-specific adaptation. This formulation has two key advantages:
\begin{itemize}
    \item Aggressive shrinkage of noise: For small or irrelevant coefficients, the prior places substantial mass near zero, effectively shrinking these estimates toward zero and helping reduce overfitting.
    \item 
    Heavy-tailed robustness: The heavy tails of the half-Cauchy distribution ensure that large signals (i.e., truly nonzero coefficients) are not overly shrunk, allowing for better recovery of strong effects.
\end{itemize}
This combination of strong shrinkage for small coefficients and minimal shrinkage for large ones yields a behavior analogous to spike-and-slab priors, but with a fully continuous and computationally tractable formulation. Moreover, the Horseshoe prior has been shown to possess optimal posterior concentration properties in sparse settings (e.g., \citealp{van2017adaptive}), making it an appealing choice both in theory and practice.

\section{Posterior inference via Gibbs Sampler}
\label{sc_gibbs_sampler}

To perform posterior inference, we implement a Gibbs sampler combining the Polya-Gamma augmentation framework from \cite{polson2013bayesian} to handle the logistic link in $\mu_i = \text{logit}^{-1}(\eta)$, and a hierachical trick for Horseshoe prior as presented in \cite{makalic2015simple}.

We observe $y_i \in (0, 1)$ and covariates $X_i \in \mathbb{R}^p$, for $i = 1, \dots, n$. The model, with given $\phi $, is:

$$
y_i \sim \text{Beta}(\mu_i \phi, (1 - \mu_i)\phi), \quad \mu_i = \frac{1}{1 + e^{-\eta}}, \quad \eta =X_i^\top \beta
.
$$
Employing the hierachical approach described in \cite{makalic2015simple}, the Horseshoe prior for the coefficients can be expressed as:
\[
\begin{aligned}
\beta_j \mid \lambda_j, \tau &\sim \mathcal{N}(0, \tau^2 \lambda_j^2),
\quad j = 1, \ldots , p,  
\\
\lambda_j^2 \mid \nu_j &\sim \text{Inv-Gamma}\left(\frac{1}{2}, \frac{1}{\nu_j} \right), 
\quad 
\nu_j \sim \text{Inv-Gamma}\left(\frac{1}{2}, 1 \right) ,
\\
\tau^2 \mid \xi &\sim \text{Inv-Gamma}\left(\frac{1}{2}, \frac{1}{\xi} \right), 
\quad 
\xi \sim \text{Inv-Gamma}\left(\frac{1}{2}, 1 \right)
.
\end{aligned}
\]
The likelihood for the data given $\beta $ is:
$$
L(  \mathbf{y} \mid \beta ) = \prod_{i=1}^n \frac{\Gamma(\phi)}{\Gamma(\mu_i \phi)\Gamma((1 - \mu_i)\phi)} y_i^{\mu_i \phi - 1}(1 - y_i)^{(1 - \mu_i)\phi - 1}.
$$
Gibbs sampling procedure: At each step, update the parameters as follows,
\begin{itemize}
\item Step 1. Sample latent Polya-Gamma variables $\omega\mid \cdot $.

Introduce $\omega\sim \text{Polya-Gamma}(\phi, \eta)$ using the identity:
$$
\frac{(e^{\eta})^{y_i \phi}}{(1 + e^{\eta})^{\phi}} 
\propto 
\exp\left( \eta \kappa_i \right) \int_0^\infty \exp\left(-\frac{\omega\eta^2}{2}\right) p(\omega)\,d\omega,
$$
with $\kappa_i = \phi \left(y_i - \frac{1}{2}\right)$. Then:
$$
\omega\sim \text{Polya-Gamma}(\phi, \eta).
$$

\item Step 2. Sample $\beta \mid \omega, \lambda^2, \tau^2$.

From the data augmentation using Polya-Gamma variables \cite{polson2013bayesian}, we can rewrite the Beta log-likelihood (in the logit link form) as:
$$
\log p(y_i \mid \eta ) 
\propto 
\kappa_i \eta - \frac{\omega\eta^2}{2}, \quad \text{where } \kappa_i = \phi(y_i - 0.5).
$$
Let $\boldsymbol{\kappa} = [\kappa_1, \ldots, \kappa_n]^\top$ 
and with $ \eta =X \beta$, 
and define $ \Omega = \mathrm{diag}(\omega_1, \ldots, \omega_n)$.
Then the augmented likelihood becomes:
$$
p(\mathbf{y} \mid \beta, \omega) \propto \exp\left(-\frac{1}{2} \beta^\top X^\top \Omega X \beta + \beta^\top X^\top \boldsymbol{\kappa} \right).
$$
Under the Horseshoe prior, the prior is:
$$
\beta \sim \mathcal{N}(0, \Lambda \tau^2), \quad \text{where } \Lambda = \mathrm{diag}(\lambda_1^2, \dots, \lambda_p^2).
$$
So the prior density is:
$$
p(\beta) \propto \exp\left( -\frac{1}{2} \beta^\top (\Lambda \tau^2)^{-1} \beta \right).
$$
Combining the likelihood and prior (both Gaussian in $\beta$), we get the full conditional Posterior for $ \beta$:
$$
p(\beta \mid \_ ) \propto \exp\left( -\frac{1}{2} \beta^\top \left(\mathbf{X}^\top \Omega \mathbf{X} + \Lambda^{-1} \tau^{-2} \right) \beta + \beta^\top X^\top \boldsymbol{\kappa} \right),
$$
which is the kernel of a multivariate normal distribution.
Thus, the conditional posterior is multivariate normal:
$$
\beta \sim \mathcal{N}(\mathbf{m}, \mathbf{V}),
$$
where
$$
\mathbf{V} = \left(X^\top \Omega \mathbf{X} + \Lambda^{-1} \right)^{-1}, \quad \mathbf{m} = \mathbf{V}X^\top \boldsymbol{\kappa},
$$
$$
\Omega= \text{diag}(\omega_1, \dots, \omega_n), 
\quad 
\Lambda^{-1} = \text{diag}\bigg( \frac{1}{\lambda_j^2 \tau^2} \bigg), 
\quad 
\kappa_i = \phi(y_i - 0.5).
$$

\item Step 3. Sample local shrinkage variances $\lambda_j^2 \mid \beta_j, \tau^2, \nu_j$.

Each $\lambda_j^2$ has a full conditional:
\[
p(\lambda_j^2 \mid \beta_j, \tau, \nu_j ) 
\propto 
\mathcal{N}(\beta_j \mid 0, \tau^2 \lambda_j^2) 
\cdot 
\text{Inv-Gamma}\left(\lambda_j^2 \mid \frac{1}{2}, \frac{1}{\nu_j} \right)
,
\]
which is conjugate and we get
$$
\lambda_j^2 \sim \text{Inverse-Gamma}\left(1, \frac{1}{\nu_j} + \frac{\beta_j^2}{2\tau^2} \right).
$$

\item Step 4. Sample local hyperparameters $\nu_j \mid \lambda_j^2$.
$$
\nu_j \sim \text{Inverse-Gamma}\left(1, 1 + \frac{1}{\lambda_j^2} \right).
$$

\item Step 5. Sample global shrinkage variance $\tau^2 \mid \beta, \lambda^2, \xi$.

We have that
\[
p(\tau^2 \mid \beta, \lambda, \xi ) 
\propto 
\prod_{j=1}^p \mathcal{N}(\beta_j \mid 0, \tau^2 \lambda_j^2) 
\cdot 
\text{Inv-Gamma} \left(\tau^2 \mid \frac{1}{2}, \frac{1}{\xi} \right)
.
\]
and thus
$$
\tau^2 \sim \text{Inverse-Gamma}\left( \frac{p + 1}{2}, \frac{1}{\xi} + \frac{1}{2} \sum_{j=1}^p \frac{\beta_j^2}{\lambda_j^2} \right).
$$

\item Step 6. Sample global hyperparameter $\xi \mid \tau^2$.
$$
\xi \sim \text{Inverse-Gamma}\left(1, 1 + \frac{1}{\tau^2} \right).
$$
\end{itemize}

After discarding the burn-in samples, use the posterior draws to estimate quantities of interest (e.g., posterior mean of $\beta$, credible intervals). Our method is implemented in the \texttt{R} package ``\texttt{betaregbayes}" available on Github: \url{https://github.com/tienmt/betaregbayes}.

\section{Theoretical result}
\label{sc_theory}
Let $\alpha\in(0,1)$ and $P,R$ be two probability measures. The $\alpha$-R\'enyi divergence  between two probability distributions $P$ and $R$ is  defined by
\begin{align*}
D_{\alpha}(P,R)  =
\frac{1}{\alpha-1} \log \int 
({\rm d}P )^\alpha 
({\rm d}R )^{1-\alpha} ,
\end{align*}
and the Kullback-Leibler divergence is defined by
$
 KL(P,R)  = 
\int \log \left(\frac{{\rm d}P}{{\rm d}R} \right){\rm d}P $  if  $ P \ll R
$, and  $
+ \infty$ otherwise.

Hereafter, we formulate some required conditions for our theoretical analysis.

\begin{assume}
\label{assum_grow_p}
    It is assumed that  \( p\) can grow at most as \( \exp(n^b) \) for some \( b < 1 \).
\end{assume}

\begin{assume}
\label{assum_beta0_bounded}
    Assume that there exists a positive constant $C_1 $ such that $ \|\beta_0\|_\infty \leq C_1  $.
\end{assume}

\begin{assume}
\label{asmum_random_design}
Assume that there exists a positive constant $ C_{\rm x} $ such that  $\mathbb{E} \| X\|^2 \leq C_{\rm x} < \infty $.
\end{assume}

\begin{assume}
\label{asmum_finite_mean}
Assume that there exists a positive constant $ C_2 $ such that  $  | X^\top \beta_0 | \leq C_2 < \infty $ almost surely.
\end{assume}

Before proceeding further, it is worth discussing the role and motivation behind each of the key assumptions.
Assumption \ref{assum_grow_p} imposes a condition on the growth rate of the dimensionality $p$ relative to the sample size $n$. 
This is a widely adopted requirement in the high-dimensional statistics literature, ensuring that consistent estimation remains possible as $n \to \infty$; see, for example, \cite{castillo2015bayesian, bellec2018slope}.
Assumption \ref{assum_beta0_bounded} places a constraint on the true parameter vector $\beta_0$. 
By bounding its norm, we restrict the complexity of the underlying signal. 
This assumption serves to prevent degenerate cases and has precedent in recent theoretical work, such as \cite{mai2024concentration, chakraborty2020bayesian}.
Assumption \ref{asmum_random_design} addresses the properties of the random design matrix. 
It ensures that the covariates are suitably well-behaved, which is essential for deriving concentration results under randomness in the design. 
Similar assumptions have been employed in earlier  Bayesian concentration analyses, including \cite{alquier2020concentration}.
Finally, Assumption \ref{asmum_finite_mean} governs the behavior of the mean response $\mu$ in Beta model. 
Specifically, it prevents $\mu$ from being too close to the boundaries (0 or 1), which could otherwise compromise model stability.

We are now in a position to formally state our main theoretical results, which establish the posterior consistency and derive concentration rates under the assumptions discussed. These results are summarized in the following theorem.

\begin{theorem}
\label{theorem_result_dis_expectation}
	For any $\alpha\in(0,1)$, 
assume that Assumption \ref{assum_grow_p}, \ref{assum_beta0_bounded}, \ref{asmum_random_design} and \ref{asmum_finite_mean} hold. We have that
\begin{equation}
\label{eq_consistency}
  \mathbb{E} 
\left[ \int D_{\alpha}(P_{\beta},P_{\beta_0}) \pi_{n,\alpha}({\rm d}\beta ) \right]
\leq \frac{1+\alpha}{1-\alpha}\varepsilon_n
	,
\end{equation}
    \begin{equation}
 \mathbb{P}\left[
\int D_{\alpha}(P_{\beta},P_{\beta_0}) \pi_{n,\alpha}({\rm d}\beta )  
\leq 
\frac{2(\alpha+1)}{1-\alpha} \varepsilon_n\right] 
  \geq 
1-\frac{2}{n\varepsilon_n}
,
\label{eq_concentration}
\end{equation}
 where
 $
 \varepsilon_n
 =
K   s^* \log \left( p /s^*\right) / n
 $, for some numerical constant $ K>0 $ depending only on $C_1, C_2, C_{\rm x} $.
  \end{theorem}

We direct the reader to \ref{sc_proofs} for all detailed technical proofs. In Theorem \ref{theorem_result_dis_expectation}, we present the main theoretical guarantees underpinning our approach. These results build primarily on foundational work concerning fractional posteriors, as developed in \cite{bhattacharya2016bayesian,alquier2020concentration}. For recent advances and applications of fractional posteriors, the reader may refer to \cite{mai2024concentration,mai2024gbayslogistic,mai2024properties,mai2025concentration}.

More particularly, the results from Theorem \ref{theorem_result_dis_expectation} show that:
\\
i) Inequality \eqref{eq_consistency} tells us that the fractional posterior is consistent when measured by the $\alpha$-R'enyi divergence. As the amount of data increases (in accordance with Assumption \ref{assum_grow_p}), the posterior distribution of $\beta$ progressively tightens around the true parameter value $\beta_0$.
\\
ii) Inequality \eqref{eq_concentration} takes this a step further by giving us the actual rate, $\varepsilon_n$, at which this concentration happens. 
This means not only do we learn the true parameter in the long run, but we also get a handle on how quickly that learning occurs — showing the fractional posterior works efficiently even in high-dimensional, sparse settings. 
Moreover, the probability bound
$
1 - \frac{2}{K s^* \log\left(p / s^*\right)}
$
remains valid and meaningful, even when the number of parameters $p$ exceeds the sample size $n$.
\\
All in all, these results provide strong theoretical backing that applies well to high-dimensional sparse Beta regression models.

Building on the findings from \cite{van2014renyi}, which explore the relationships among the Hellinger distance, total variation, and the 
$\alpha$-R\'enyi divergence, we derive the following concentration results:
$$
    \mathbb{P}\left[
\int H^2 (P_{\beta},P_{\beta_0}) \pi_{n,\alpha}({\rm d}\beta )  
\leq 
K_\alpha
\varepsilon_n\right] 
  \geq 
1-\frac{2}{n\varepsilon_n}
,
$$
$$
\mathbb{P}\left[
\int d^{2}_{TV} (P_{\beta},P_{\beta_0}) \pi_{n,\alpha}({\rm d}\beta )  
\leq 
K_\alpha \varepsilon_n\right] 
  \geq 
1-\frac{2}{n\varepsilon_n}
,
$$
where $d_{TV}$ denotes the total variation distance and $ H^2 $ represents the squared Hellinger distance and $ K_\alpha $ is a  positive  constant depending solely on $ \alpha $. Details can be found in \cite{alquier2020concentration}.

While results based on different divergences help us understand how the posterior behaves, they do not directly tell us how close the estimates are in terms of standard Euclidean distance. To address this, we now derive concentration results using the $\ell_2$ distance.

\begin{prop}
\label{propo_1}
Assuming that Theorem \ref{theorem_result_dis_expectation} holds, there exists a constant $ c>0 $ such that
\begin{equation}
  \mathbb{E} 
\left[ \int \Vert X^\top \! (\beta-  \beta_0) \Vert_2^2 \pi_{n,\alpha}({\rm d}\beta ) \right]
\leq 
\frac{1+\alpha}{(1-\alpha) c } \varepsilon_n
	,
\end{equation}
\begin{equation}
 \mathbb{P}\left[
\int \Vert X^\top \! (\beta-  \beta_0) \Vert_2^2 
\pi_{n,\alpha}({\rm d}\beta )  
\leq 
\frac{2(\alpha+1)}{ (1-\alpha) c } 
\varepsilon_n\right] 
  \geq 
1-\frac{2}{n\varepsilon_n}
.
\end{equation}
\end{prop}

A direct consequence of the above results is a bound for the Bayesian mean estimator. Let
$$
\hat{\beta} := \int \beta\, \pi_{n,\alpha}(\beta)\, \mathrm{d}\beta
$$
be our posterior mean estimator. By exploiting the convexity of the $\ell_2 $ norm, we obtain the following bound:
\begin{equation*}
  \mathbb{E} 
 \Vert X^\top \! (\hat{ \beta } -  \beta_0) \Vert_2^2 
\leq 
\frac{1+\alpha}{(1-\alpha) c } \varepsilon_n
	,
\end{equation*}
\begin{equation*}
 \mathbb{P}\left[
\Vert X^\top \! ( \hat{ \beta } -  \beta_0) \Vert_2^2   
\leq 
\frac{2(\alpha+1)}{ (1-\alpha) c } 
\varepsilon_n\right] 
  \geq 
1-\frac{2}{n\varepsilon_n}
.
\end{equation*}

\section{Numerical studies}
\label{sc_numerical}
In this section, we present numerical studies of our method on both simulated and real datasets. Since no existing method is available for direct comparison in high-dimensional sparse Beta regression, we employ the Lasso method applied to transformed responses as a benchmark. To ensure a fairer comparison, we also conduct simulation studies in low-dimensional settings, where well-established methods exist—specifically, the ``\texttt{betareg}" package in \texttt{R}, \cite{cribari2010beta}.

\subsection{Simulation in low dimensions}

We generate $ X_i \sim N(0, \Sigma)$ and consider the following covariance structures for the predictors: an independent structure with $\Sigma = \mathbb{I}_p$, and a correlated structure with entries defined by $(\Sigma)_{ij} = \rho_X^{|i-j|}$ for all $i, j$. 
The responses $y_i$ are generated from a Beta distribution as specified in equation \eqref{eq_beta_distribut}, with the precision parameter fixed at $\phi = 10$ and the mean $
 \mu_i = \text{logit}^{-1}(X_i^\top \beta_0).
$. 
To fit the Lasso model, we transform the response variable from $y$ to $y^* := \log\big(\frac{y}{1 - y}\big)$, and then apply Lasso regression to $y^*$.
In this low-dimensional setting, we set $p = 20$ and the number of non-zero coefficients $s^* = 10$. The first half of the non-zero entries in $\beta_0$ are set to 1, and the second half to -1. We vary the sample size $n \in \{100, 500, 1000\}$.

We consider the following error metrics to quantify the estimation accuracy: 
$$
\ell_2(\beta_0) : = p^{-1} \| \widehat{\beta} -\beta_0 \|_2^2 ,
\quad
\ell_2( X^\top \!\! \beta_0)
: = 
n^{-1} \| X^\top \! \widehat{\beta} - X^\top \! \! \beta_0 \|_2^2 ,
$$
in which $ \widehat{\beta} $ is the considered methods, and the following error metrics to quantify the prediction accuracy
$$
  \ell_2(Y)
  : = n^{-1} \sum_{i=1}^n 
\left( y_i - \widehat{\mu_i} \right)^2 , 
\text{ with } \, 
\widehat{\mu_i} 
=
\text{logit}^{-1}(X_i^\top \widehat{\beta} )
$$
$$
\ell_2(Y_{\rm test})
 := 
n_{\rm test}^{-1} \sum_{i=1}^{n_{\rm test}} 
\left( y_{{\rm test},i}  -
\widehat{\mu}_{{\rm test},i} 
\right)^2  ,
\text{ with } \, 
\widehat{\mu}_{{\rm test},i} 
=
\text{logit}^{-1}(X_{{\rm test}}^\top \widehat{\beta} )
.
$$
Here, $ y_{{\rm test}} $ and $ X_{{\rm test}} $ are testing data generated as $y $ (using $\beta_0 $) and we take $ n_{\rm test} =30 $ for all setting. 
In addition, we also evaluate the variable selection performance of the proposed method. 
Specifically, we compute the standard variable selection metrics based on the counts of true positives (TP), false negatives (FN), false positives (FP), and true negatives (TN). The following measures are considered:
$$
\displaystyle \text{Precision} = \frac{\text{TP}}{\text{TP} + \text{FP}}
; \quad
\displaystyle \text{Recall} = \frac{\text{TP}}{\text{TP} + \text{FN}}
; \quad
\displaystyle \text{Specificity} = \frac{\text{TN}}{\text{TN} + \text{FP}}
;
$$
$$
\displaystyle \text{F1} = \frac{2 \cdot \text{Precision} \cdot \text{Recall}}{\text{Precision} + \text{Recall}}
; \quad
\displaystyle \text{FDR} = \frac{\text{FP}}{\text{TP} + \text{FP}}
.
$$
These metrics collectively assess both the accuracy and reliability of the variable selection procedure.

\begin{table}[!ht]
\centering
\caption{Simulation results in low dimensions, with $ p = 20, s^* = 10 $.}
	\begin{tabular}{ l | ccc | ccc  }
		\hline \hline
Method 
& Betareg & Lasso   & Horseshoe 
 & Betareg & Lasso   & Horseshoe 
\\
\hline
& \multicolumn{3}{c|}{  $ n = 100 $ }
&  \multicolumn{3}{c}{ $ n = 100,  \rho_X = 0.5 $ }
\\ 
\hline
$ 10 \times \ell_2(\beta_0) $ 
& 0.10 (0.05) & 2.31 (0.63) & 0.09 (0.04)
& 0.33 (0.15) & 2.22 (0.67) & 0.21 (0.10) 
		\\
$ \ell_2( X^\top\!\! \beta_0) $
& 0.18 (0.08) & 4.69 (1.41) &  0.16 (0.06)
& 0.84 (0.57) & 6.72 (1.62) & 0.28 (0.15) 
		\\
$ 10 \times \ell_2(Y)$ 
& 0.08 (0.02) & 0.17 (0.04) & 0.07 (0.02)
& 0.07 (0.02) & 0.17 (0.05) & 0.05 (0.01) 
		\\
$ 10 \times \ell_2(Y_{\rm test}) $ 
& 0.13 (0.05) & 0.21 (0.09) & 0.13 (0.06) 
& 0.12 (0.05) & 0.20 (0.11) & 0.11 (0.05) 
\\
Precision 
& 0.92 (0.08) & 0.62 (0.08) & 0.98 (0.04)
& 0.93 (0.08) & 0.70 (0.12) & 0.99 (0.03)
\\
Recall           
& 1.00 (0.00) & 1.00 (0.00) & 1.00 (0.00)
& 1.00 (0.00) & 1.00 (0.00) & 1.00 (0.00)
\\
F1 
& 0.96 (0.05) & 0.77 (0.06) & 0.99 (0.02)
& 0.96 (0.04) & 0.81 (0.08) & 1.00 (0.02)
\\
Specificity          
& 0.90 (0.11) & 0.37 (0.20) & 0.98 (0.04)
& 0.92 (0.10) & 0.52 (0.26) & 0.99 (0.03)
\\
FDR  
& 0.08 (0.08) & 0.38 (0.08) & 0.02 (0.04)
& 0.07 (0.08) & 0.30 (0.12) & 0.01 (0.03)
		\\
		\hline
& \multicolumn{3}{c|}{ $ n = 500 $ }
&  \multicolumn{3}{c}{ $ n = 500,  \rho_X = 0.5 $ }
\\ 
\hline
$ 10 \times \ell_2(\beta_0) $ 
& 0.06 (0.02) & 2.34 (0.30) & 0.02 (0.01) 
& 0.27 (0.07) & 1.91 (0.26) & 0.03 (0.01) 
		\\
$ \ell_2( X^\top\!\! \beta_0) $
& 0.12 (0.04) & 4.66 (0.67) & 0.03 (0.01) 
& 0.96 (0.31) & 6.89 (0.82) & 0.05 (0.02) 
		\\
$ 10 \times \ell_2(Y)$ 
& 0.10 (0.01) & 0.14 (0.01) & 0.10 (0.01) 
& 0.09 (0.01) & 0.12 (0.01) & 0.07 (0.01) 
		\\
$ 10 \times \ell_2(Y_{\rm test}) $ 
& 0.10 (0.03) & 0.14 (0.05) & 0.10 (0.03) 
& 0.10 (0.03) & 0.13 (0.06) & 0.09 (0.04) 
\\
Precision 
& 0.95 (0.06) & 0.63 (0.10) & 0.98 (0.04)
& 0.97 (0.06) & 0.67 (0.11) & 0.99 (0.03)
\\
Recall           
& 1.00 (0.00) & 1.00 (0.00) & 1.00 (0.00)
& 1.00 (0.00) & 1.00 (0.00) & 1.00 (0.00)
\\
F1 
& 0.97 (0.03) & 0.77 (0.07) & 0.99 (0.02)
& 0.98 (0.03) & 0.80 (0.07) & 0.99 (0.01)
\\
Specificity          
& 0.94 (0.07) & 0.38 (0.24) & 0.98 (0.04)
& 0.96 (0.07) & 0.48 (0.23) & 0.99 (0.03)
\\
FDR  
& 0.05 (0.06) & 0.37 (0.10) & 0.02 (0.04)
& 0.03 (0.06) & 0.33 (0.11) & 0.99 (0.03)
		\\
		\hline
& \multicolumn{3}{c|}{ $ n = 1000 $ }
&  \multicolumn{3}{c}{ $ n = 1000,  \rho_X = 0.5 $ }
\\ 
\hline
$ 10^2 \times \ell_2(\beta_0) $ 
& 0.51 (0.15) & 23.3 (2.40) & 0.09 (0.04) 
& 2.59 (0.44) & 19.5 (1.99) & 0.16 (0.06) 
		\\
$ \ell_2( X^\top\!\! \beta_0) $
& 0.10 (0.03) & 4.67 (0.56) & 0.02 (0.01) 
& 0.94 (0.19) & 7.13 (0.58) & 0.03 (0.01) 
		\\
$ 10 \times \ell_2(Y)$ 
& 0.10 (0.01) & 0.14 (0.01) & 0.10 (0.01) 
& 0.09 (0.01) & 0.11 (0.01) & 0.07 (0.01) 
		\\
$ 10 \times \ell_2(Y_{\rm test}) $ 
& 0.11 (0.04) & 0.14 (0.06) & 0.11 (0.04) 
& 0.09 (0.03) & 0.11 (0.05) & 0.08 (0.03) 
\\
Precision 
& 0.95 (0.06) & 0.62 (0.09) & 0.98 (0.04)
& 0.95 (0.08) & 0.70 (0.11) & 0.99 (0.03)
\\
Recall           
& 1.00 (0.00) & 1.00 (0.00) & 1.00 (0.00)
& 1.00 (0.00)  & 1.00 (0.00)  & 1.00 (0.00) 
\\
F1 
& 0.98 (0.03) & 0.76 (0.07) & 0.99 (0.02)
& 0.97 (0.05) & 0.82 (0.07) & 0.99 (0.02)
\\
Specificity          
& 0.95 (0.07) & 0.36 (0.23) & 0.98 (0.04)
& 0.94 (0.15) & 0.54 (0.22) & 0.98 (0.04)
\\
FDR  
& 0.05 (0.06) & 0.38 (0.09) & 0.02 (0.04)
& 0.05 (0.08) & 0.30 (0.11) & 0.01 (0.03)
		\\
		\hline
		\hline	
\end{tabular}
\label{tb_low_dim}
\end{table}

We run the Gibbs sampler for 1200 iterations, discarding the first 200 as burn-in, and fix $\alpha = 0.99$. Our proposed method is referred to as “Horseshoe”. 
For the Lasso, we use the default settings and perform 10-fold cross-validation to select the optimal tuning parameter. 
For the maximum likelihood estimation using Beta regression, we also use default options as implemented in the ``\texttt{betareg}" package, and refer to this method as “Betareg”. 
For each simulation setting, we generate 100 independent datasets and report the average results along with their standard deviations. The results are given in Table \ref{tb_low_dim}.

The results presented in Table \ref{tb_low_dim} demonstrate that, in settings with a large sample size, our Horseshoe-based method outperforms both Beta regression and the transformed Lasso approach. Notably, the suboptimal performance of the transformed Lasso can be attributed to its reliance on linear model assumptions, which are not well-suited to this context. 
Compared to Betareg method, which represents the state-of-the-art maximum likelihood approach in low-dimensional settings, our method achieves up to a fourfold reduction in estimation error. While the improvement in prediction error is more modest, our method still performs slightly better. 
With respect to variable selection, our Horseshoe method demonstrates consistently superior accuracy, outperforming both Betareg method and the transformed Lasso methods in all considered settings.
These findings provide strong evidence that our approach delivers highly accurate results, even in traditional, well-studied scenarios.

\subsection{Simulation in high dimensions}

\begin{table}[!ht]
\centering
\caption{Simulation results for high dimensional settings.}
	\begin{tabular}{ l | cc | cc  }
		\hline \hline
Method 
 & Lasso   & Horseshoe 
  & Lasso   & Horseshoe 
\\
\hline
$ n = 80,p=100$ 
& \multicolumn{2}{c|}{  $ s^* = 10 $ }
&  \multicolumn{2}{c}{ $  s^* = 10 ,  \rho_X = 0.5 $ }
\\ 
\hline
$ 10 \times \ell_2(\beta_0) $ 
& 0.37 (0.19) & 0.03 (0.02) 
& 0.50 (0.19) & 0.15 (0.12) 
		\\
$ \ell_2( X^\top\!\! \beta_0) $
& 4.28 (1.80) & 0.29 (0.13)  
& 6.39 (2.28) & 0.68 (0.42) 
		\\
$ 10 \times \ell_2(Y)$ 
& 0.20 (0.07) & 0.03 (0.01) 
& 0.19 (0.07) & 0.02 (0.01) 
		\\
$ 10 \times \ell_2(Y_{\rm test}) $ 
& 0.35 (0.22) & 0.16 (0.06) 
& 0.26 (0.13) & 0.19 (0.10) 
		\\
Precision 
& 0.31 (0.08) & 1.00 (0.00)
& 0.42 (0.13) & 1.00 (0.00)
\\
Recall           
& 1.00 (0.01) &  0.99 (0.03)
& 0.99 (0.04) &  0.72 (0.15)
\\
F1 
& 0.46 (0.09) & 1.00 (0.02)
& 0.58 (0.12) & 0.83 (0.10)
\\
Specificity          
& 0.72 (0.11) &  1.00 (0.00)
& 0.82 (0.08) &  1.00 (0.00)
\\
FDR  
& 0.69 (0.08) & 0.00 (0.00)
& 0.58 (0.13) &  0.00 (0.00)
\\
		\hline
$ n = 80,p=100$  
& \multicolumn{2}{c|}{ $ s^* = 20 $ }
&  \multicolumn{2}{c}{ $ s^* = 20,  \rho_X = 0.5 $ }
\\ 
\hline
$ 10 \times \ell_2(\beta_0) $ 
& 0.96 (0.35) & 0.65 (0.20) 
& 1.22 (0.58) &  0.90 (0.30)
		\\
$ \ell_2( X^\top\!\! \beta_0) $
& 8.16 (2.14) & 2.16 (0.66) 
& 6.36 (3.01) & 3.91 (1.84) 
		\\
$ 10 \times \ell_2(Y)$ 
& 0.23 (0.12) &  0.01 (0.00) 
& 0.26 (0.11) &  0.00 (0.00)
		\\
$ 10 \times \ell_2(Y_{\rm test}) $ 
& 0.72 (0.39) &  0.80 (0.37)
& 0.59 (0.34) & 0.53 (0.31) 
		\\
Precision 
& 0.40 (0.07) & 0.98 (0.07)
& 0.57 (0.12) & 1.00 (0.00)
\\
Recall           
& 0.99 (0.03) &  0.43 (0.13)
& 0.91 (0.08) &  0.21 (0.08)
\\
F1 
& 0.57 (0.07) & 0.59 (0.13)
& 0.69 (0.09) & 0.34 (0.11)
\\
Specificity          
& 0.62 (0.11) &  1.00 (0.01)
& 0.80 (0.10) &  1.00 (0.00)
\\
FDR  
& 0.60 (0.07) & (0.02) (0.07)
& 0.43 (0.12) &  0.00 (0.00)
\\
		\hline
$ n=200, p = 300 $ & \multicolumn{2}{c|}{ $ s^* = 10 $ }
&  \multicolumn{2}{c}{ $ s^* = 10,  \rho_X = 0.5 $ }
\\ 
\hline
$ 10^2 \times \ell_2(\beta_0) $ 
& 0.87 (0.30) & 0.03 (0.01) 
& 1.01 (0.26) &  0.05 (0.02)
		\\
$ \ell_2( X^\top\!\! \beta_0) $
& 3.16 (1.15) & 0.09 (0.04) 
& 5.13 (1.35) &  0.13 (0.07)
		\\
$ 10 \times \ell_2(Y)$ 
& 0.14 (0.03) &  0.05 (0.01)
& 0.15 (0.04) & 0.04 (0.01) 
		\\
$ 10 \times \ell_2(Y_{\rm test}) $ 
& 0.17 (0.08) &  0.11 (0.05)
& 0.17 (0.09) & 0.09 (0.04) 
		\\
Precision 
& 0.25 (0.08) & 1.00 (0.01)
& 0.33 (0.13) & 1.00 (0.00)
\\
Recall           
& 1.00 (0.00) &  1.00 (0.00)
& 1.00 (0.00) &  1.00 (0.00)
\\
F1 
& 0.39 (0.11) & 1.00 (0.00)
& 0.49 (0.14) & 1.00 (0.00)
\\
Specificity          
& 0.87 (0.07) &  1.00 (0.00)
& 0.92 (0.05)  &  1.00 (0.00)
\\
FDR  
& 0.75 (0.08) & 0.00 (0.00)
& 0.67 (0.13) &  0.00 (0.00)
\\
		\hline
$ n=200, p = 300 $ & \multicolumn{2}{c|}{ $ s^* = 20 $ }
&  \multicolumn{2}{c}{ $ s^* = 20,  \rho_X = 0.5 $ }
\\ 
\hline
$ 10^2 \times \ell_2(\beta_0) $ 
& 1.35 (0.37) &  0.09 (0.03)
& 1.50 (0.51) &  0.76 (0.58)
		\\
$ \ell_2( X^\top\!\! \beta_0) $
& 5.11 (1.34) & 0.28 (0.09) 
& 3.46 (1.21) &  1.06 (0.64)
		\\
$ 10 \times \ell_2(Y)$ 
& 0.16 (0.04) & 0.02 (0.00) 
& 0.19 (0.05) &  0.01 (0.00)
		\\
$ 10 \times \ell_2(Y_{\rm test}) $ 
& 0.25 (0.12) & 0.11 (0.05) 
& 0.26 (0.12) & 0.16 (0.10) 
		\\
Precision 
& 0.28 (0.05) & 1.00 (0.00)
& 0.43 (0.11) & 1.00 (0.00)
\\
Recall           
& 1.00 (0.00) &  1.00 (0.00)
& 0.99 (0.02) &  0.80 (0.14)
\\
F1 
& 0.43 (0.06) & 1.00 (0.00)
& 0.59 (0.10) & 0.88 (0.09)
\\
Specificity          
& 0.81 (0.05) &  1.00 (0.00)
& 0.89 (0.05) &  1.00 (0.00)
\\
FDR  
& 0.72 (0.05) & 0.00 (0.00)
& 0.57 (0.11) &  0.00 (0.00)
\\
		\hline
		\hline	
\end{tabular}
\label{tb_highdim}
\end{table}

In this section, the simulations are conducted in the same manner as before; however, we now focus on scenarios where the number of predictors significantly exceeds the sample size. Specifically, we consider two settings: $n = 80, p = 100$ and $n = 200, p = 300$. We also vary the sparsity levels by setting $s^* \in \{10, 20\}$. Since the \texttt{betareg} package requires $n > p$, it is not applicable in this high-dimensional setting. Therefore, we compare our method only with the transformed Lasso approach.

The results shown in Table \ref{tb_highdim} consistently demonstrate that our Horseshoe-based method performs effectively in high-dimensional settings, yielding notably low estimation errors. 
As anticipated, the transformed Lasso continues to perform poorly in this context, and our method achieves substantially lower prediction errors in comparison. 
Once again, in the context of variable selection, our proposed method demonstrates outstanding performance. Across a wide range of high-dimensional scenarios, it consistently identifies the true set of predictors with remarkable accuracy. This robustness in selection highlights the method’s ability to effectively distinguish relevant variables from noise, even in challenging settings where the number of predictors greatly exceeds the number of observations.
This series of simulation studies provides strong empirical support for the theoretical properties of our proposed approach.

To further assess the behavior of the Gibbs sampler in high-dimensional settings, we provide trace plots and autocorrelation function (ACF) plots in the Appendix (Figures \ref{fig_tracplot} and \ref{fig_acf_plot}). These diagnostics offer insights into the convergence and mixing properties of the sampler.

\subsection{Simulation results for changing the precision $ \phi $}
Since our method assumes a fixed precision parameter $\phi$ in the Beta regression model, 
we now conducte a series of simulations to assess the sensitivity of the method to the choice of this parameter. 
The simulation settings are identical to those described previously.
We consider both a low-dimensional scenario with $n = 100, p = 20, s^* = 10$, and a high-dimensional scenario with $n = 80, p = 100, s^* = 10$. In all cases, the true value of $\phi$ is set to 10. 
Each simulation setup is repeated 100 times, and Table \ref{tb_change_phi} reports the average and standard deviation of the results.

\begin{table}[!ht]
\centering
\caption{Simulation results for changing $\phi$. The true value of $\phi$ is 10 and $s^*$ is fixed at 10.}
	\begin{tabular}{ l  cc  cc c  }
		\hline \hline
Fitted value 
 & $ \phi = 1 $   & $ \phi = 5 $ 
  & $ \phi = 10 $   & $ \phi = 15 $ & $ \phi = 20 $
\\
\hline
  \multicolumn{2}{l}{ \quad\quad $ n = 100,p=20 $}
\\ 
\hline
$ 10 \times \ell_2(\beta_0) $ 
& 1.32 (0.39) & 0.10 (0.04)
& 0.09 (0.04) & 0.11 (0.06)
& 0.12 (0.05)
		\\
$ \ell_2( X^\top\!\! \beta_0) $
& 2.24 (0.65) & 0.17 (0.08)
& 0.16 (0.06)  &  0.19 (0.10)
& 0.20 (0.08)
		\\
$ 10 \times \ell_2(Y)$ 
& 0.19 (0.06) & 0.07 (0.02)
& 0.07 (0.02) &  0.07 (0.02)
& 0.07 (0.02)
		\\
$ 10 \times \ell_2(Y_{\rm test}) $ 
& 0.31 (0.11) & 0.12 (0.06)
& 0.13 (0.06) &  0.13 (0.06)
& 0.14 (0.06)
		\\
Precision 
& 1.00 (0.00) & 1.00 (0.01) & 0.99 (0.03)  & 0.96 (0.07) & 0.93 (0.06)
\\
Recall           
& 0.28 (0.16) & 1.00 (0.00) & 1.00 (0.00) & 1.00 (0.00) & 1.00 (0.00) 
\\
F1 
& 0.44 (0.17) & 1.00 (0.01) & 0.99 (0.02) & 0.98 (0.04) & 0.97 (0.03)
\\
Specificity          
& 1.00 (0.00) & 1.00 (0.01) & 0.99 (0.04) & 0.95 (0.08) & 0.92 (0.07)
\\
FDR  
& 0.00 (0.00)  & 0.00 (0.01)  & 0.01 (0.03) & 0.04 (0.07) & 0.07 (0.06)
		\\
		\hline
  \multicolumn{3}{l}{\quad\quad $ n = 100,p=20, \rho_X = 0.5$}
&
\\ 
\hline
$ 10 \times \ell_2(\beta_0) $ 
& 1.04 (0.26) & 0.23 (0.14)
& 0.21 (0.10) &  0.23 (0.12)
& 0.24 (0.13)
		\\
$ \ell_2( X^\top\!\! \beta_0) $
& 2.41 (0.86) & 0.32 (0.20)
& 0.28 (0.15) & 0.31 (0.17) 
& 0.32 (0.23)
		\\
$ 10 \times \ell_2(Y)$ 
& 0.11 (0.03) & 0.05 (0.01)
& 0.05 (0.01) & 0.05 (0.01) 
& 0.05 (0.01)
		\\
$ 10 \times \ell_2(Y_{\rm test}) $ 
& 0.20 (0.08) & 0.11 (0.06)
& 0.11 (0.05) &  0.11 (0.05)
& 0.11 (0.05)
		\\
Precision 
& 1.00 (0.00) & 1.00 (0.01) & 0.99 (0.03) & 0.97 (0.05) & 0.95 (0.07)
\\
Recall           
& 0.11 (0.08) & 0.99 (0.04) & 1.00 (0.00) & 1.00 (0.00) & 1.00 (0.00) 
\\
F1 
& 0.25 (0.09) & 0.99 (0.02) & 1.00 (0.02) & 0.99 (0.03) & 0.97 (0.04)
\\
Specificity          
& 1.00 (0.00) & 1.00 (0.01) & 0.99 (0.03) & 0.97 (0.06) & 0.94 (0.10)
\\
FDR  
& 0.00 (0.00)  & 0.00 (0.01)  & 0.01 (0.03) & 0.03 (0.05) & 0.05 (0.07)
		\\
		\hline
  \multicolumn{2}{l}{ \quad\quad $ n = 80,p=100 $}\\ 
\hline
$ 10^2 \times \ell_2(\beta_0) $ 
& 0.85 (0.09) & 0.07 (0.04)
& 0.03 (0.02) &  0.04 (0.02)
& 0.05 (0.02)
		\\
$ \ell_2( X^\top\!\! \beta_0) $
& 7.80 (1.33) & 0.49 (0.29)
& 0.29 (0.13) &  0.33 (0.13)
& 0.43 (0.19)
		\\
$ 10 \times \ell_2(Y)$ 
& 0.92 (0.23) & 0.04 (0.01)
& 0.03 (0.01) &  0.02 (0.01)
& 0.02 (0.00)
		\\
$ 10 \times \ell_2(Y_{\rm test}) $ 
& 1.19 (0.22) & 0.21 (0.10)
& 0.16 (0.06) &   0.16 (0.06)
& 0.17 (0.08)
		\\
Precision 
& 1.00 (0.00) & 1.00 (0.00) & 1.00 (0.00) & 1.00 (0.02) & 0.99 (0.02)
\\
Recall           
& 0.00 (0.02)  & 0.92 (0.10) & 0.99 (0.03) & 1.00 (0.01) & 1.00 (0.00) 
\\
F1 
& 0.18 (0.00) & 0.95 (0.06) & 1.00 (0.02) & 1.00 (0.01) & 1.00 (0.01)
\\
Specificity          
& 1.00 (0.00) & 1.00 (0.00) & 1.00 (0.00) & 1.00 (0.00) & 1.00 (0.00) 
\\
FDR  
& 0.00 (0.00)  & 0.00 (0.00)  & 0.00 (0.00)  & 0.00 (0.02)  & 0.01 (0.02)
		\\
		\hline
  \multicolumn{3}{l}{\quad\quad $ n = 80,p=100, \rho_X = 0.5$}\\ 
\hline
$ 10^2 \times \ell_2(\beta_0) $ 
& 0.59 (0.10) & 0.28 (0.12)
& 0.15 (0.12) &  0.14 (0.10)
& 0.11 (0.07)
		\\
$ \ell_2( X^\top\!\! \beta_0) $
& 7.42 (1.64) & 1.21 (0.50)
& 0.68 (0.42) &  0.63 (0.31)
& 0.70 (0.34)
		\\
$ 10 \times \ell_2(Y)$ 
& 0.34 (0.14) &  0.04 (0.02)
& 0.02 (0.01) &  0.01 (0.00)
& 0.01 (0.00)
		\\
$ 10 \times \ell_2(Y_{\rm test}) $ 
& 0.59 (0.21) & 0.30 (0.17)
& 0.19 (0.10) & 0.19 (0.11) 
& 0.18 (0.09)
		\\
Precision 
& 1.00 (0.00) & 1.00 (0.00) & 1.00 (0.00) & 1.00 (0.02) & 1.00 (0.02) 
\\
Recall           
& 0.03 (0.05) & 0.42 (0.12) & 0.73 (0.16) & 0.86 (0.15) & 0.88 (0.13)
\\
F1 
& 0.20 (0.05) & 0.59 (0.13) & 0.84 (0.11) & 0.92 (0.10) & 0.93 (0.08)
\\
Specificity          
& 1.00 (0.00) & 1.00 (0.00) & 1.00 (0.00) & 1.00 (0.00) & 1.00 (0.00) 
\\
FDR  
& 0.00 (0.00)  & 0.00 (0.00)  & 0.00 (0.00)  & 0.00 (0.02)  & 0.00 (0.02) 
		\\
		\hline
		\hline	
\end{tabular}
\label{tb_change_phi}
\end{table}

Table \ref{tb_change_phi} shows that, as expected, setting $\phi$ too far from its true value can negatively impact performance. However, when $\phi$ is chosen within a reasonable range around the true value, the results remain highly competitive—nearly matching those achieved using the oracle value. This suggests that, in practice, a consistent estimate of $\phi$, such as one obtained via maximum likelihood based on the response variable alone, can be effectively used in our method without significant loss in accuracy.

\subsection{Application: analyzing and predicting Student's GPA Ratio}

In this section, we demonstrate the practical utility of our proposed method through an application to a real dataset. The aim of this case study is to investigate how various aspects of the study environment are associated with academic performance, measured via students' GPA ratios. Specifically, we have the response as the ratio of each student’s GPA to the maximum possible GPA—a continuous response variable bounded in the range $ (0, 1) $, \citep{vanni2025realdata}. The dataset is publicly available at \url{https://doi.org/10.5281/zenodo.15423017}.

The dataset comprises responses from 171 university students, primarily located in Milan and Rome, Italy. 
The covariates includes both categorical and continuous variables capturing demographic characteristics and study environment conditions. The variables are as follows:
Gender;
City (Milan, Rome, Other);
Major;
Study time (Morning, Afternoon, Evening, Night);
Study location (Home, Library, Café, Other);
Age;
Enough sleep;
Natural sun exposure;
Noisy environment;
Adequate heating/cooling;
Well-ventilated;
Enough desk space;
Often distracted; 
Study in group.
This dataset enables the examination of how physical and environmental study conditions relate to students’ academic outcomes. 

Using the Betareg method, we identified several covariates associated with the GPA ratio. Specifically, ``age", ``city" (Iowa and Pavia), ``major" (Economics, Engineering, Mathematics, Medicine), ``often distracted", and ``enough desk space" were all negatively associated with GPA ratio, while good ``ventilation" showed a positive association.

In contrast, our Horseshoe regression model identified only one strong signal: the covariate ``often distracted" exhibited a significant negative effect on GPA ratio. The estimated effect under the Horseshoe model was $-0.373$ with a 95\% credible interval of $[-0.567, -0.173]$,
while the Beta regression produced a comparable estimate of $ -0.412$ with a 95\% confidence interval of $[-0.622, -0.203]$. Notably, both models provide consistent evidence for the adverse impact of frequent distraction on academic performance.

We next evaluate the predictive performance of the two methods on this dataset. 
To this end, we randomly select 35 observations (approximately 20\% of the full dataset of 171 samples) to serve as a test set, 
while the remaining 80\% is used for model fitting. 
This procedure is repeated 100 times to account for variability due to data splitting, and we report the average predictive performance across these replications in Table \ref{tb_realdata}.

\begin{table}[!h]
	\centering
	\caption{Mean (and standard deviation)  prediction errors for the real data.}
	\begin{tabular}{  l | cc }
		\hline \hline
 & Horseshoe & Betareg 	
 \\ 		\hline
$ 100\times \ell_2(Y_{\rm test}) $ & 0.453 (0.110) & 0.564 (0.165)
\\
		\hline
		\hline
	\end{tabular}
	\label{tb_realdata}
\end{table}

The results presented in Table \ref{tb_realdata} clearly indicate that our Horseshoe method outperforms the Beta regression approach in terms of predictive accuracy. This finding further supports our theoretical results and aligns with the simulation studies discussed earlier.

\section{Discussion and Conclusion}
In this paper, we developed a novel Bayesian framework for sparse beta regression tailored to high-dimensional settings. Our approach addresses a critical gap in the literature, where bounded outcomes are common but theoretical and computational tools for handling them in modern, high-dimensional contexts remain underdeveloped. Our methodology inherits strong theoretical properties, including posterior consistency and convergence rates—results that, to the best of our knowledge, are the first of their kind for Bayesian beta regression.

The integration of the Horseshoe prior into the beta regression model enables principled variable selection, yielding both interpretability and robustness in sparse settings. Importantly, we move beyond traditional Metropolis–Hastings schemes by introducing a new Gibbs sampling algorithm, leveraging Pólya–Gamma data augmentation to facilitate efficient inference. 
Our simulation and real data application results demonstrate the superior performance of our method relative to existing alternatives, in both low- and high-dimensional scenarios. 
Taken together, our contributions advance the state of the art in modeling bounded data, offering a flexible and computationally tractable approach that is suitable for a wide range of applications across scientific domains.

Potential directions for future research include incorporating structured sparsity, developing nonparametric extensions, and designing scalable variational inference methods specifically adapted to beta regression. Additionally, given the widespread occurrence of proportion data in applied settings, our framework can be extended to accommodate zero- and/or one-inflated responses, as discussed in \cite{zhou2022bayesian,liu2018review}, or adapted for robust regression in the presence of outliers, following approaches such as \cite{bayes2012new,migliorati2018new}.

\subsection*{Acknowledgments}
The views, results, and opinions presented in this paper are solely those of the author and do not, in any form, represent those of the Norwegian Institute of Public Health.

\subsection*{Conflicts of interest/Competing interests}
The author declares no potential conflict of interests.

\clearpage
\appendix
\section{Proofs}
\label{sc_proofs}

\subsection{Lemma}
We first state some important results that will be used in our proof.

\begin{lemma}
\label{lm_bounding_KL}
Let \( \mu, \mu' \in (0,1) \), \( \phi > 0 \), 
there exist some constant $ C >0$ such that for small \( \delta = \mu - \mu' \) and that $\mu \in [\epsilon, 1 - \epsilon]$, we have
$$
 KL \left( {\rm Beta} (\mu\phi, (1 - \mu)\phi) \, , \, {\rm Beta}(\mu'\phi, (1 - \mu')\phi)\right) 
\leq 
C   (\mu - \mu')^2,
$$
\end{lemma}

\begin{lemma}
\label{lm_lowerbound_reymi}
Let \( \mu, \mu' \in (0,1) \), \( \phi > 0 \), and consider the Beta distributions
\[
P = \mathrm{Beta}(\mu \phi, (1 - \mu)\phi), \quad Q = \mathrm{Beta}(\mu' \phi, (1 - \mu')\phi).
\]
There exist some constant $c>0$ such that for small \( \delta = \mu - \mu' \) and that $\mu \in [\epsilon, 1 - \epsilon]$, we have
\[
D_\alpha(P , Q) 
\geq 
c (\mu - \mu')^2 
.
\]
\end{lemma}

\begin{lemma}
\label{lm_boundloglikelihood}
There exist a positive constant $ C $ such that, for $\mu \in [\epsilon, 1 - \epsilon]$,
\begin{equation*}
    \left| \log L(\beta \mid \mathbf{y},X, \phi ) - \log L(\beta' \mid \mathbf{y},X, \phi ) \right|
\leq
C |X^\top\! \beta -X^\top\! \beta' |
    .
\end{equation*}
\end{lemma}

Proofs of Lemma 
\ref{lm_bounding_KL}, 
\ref{lm_lowerbound_reymi}, and
\ref{lm_boundloglikelihood}
 are given at the end of the proof section, \ref{sc_proof_of_lemmaaaa}.

\begin{theorem}[Theorem 2.6 in \cite{alquier2020concentration}]
\label{thm_expect_alquier}
Assume that there exists a sequence $\varepsilon_n > 0$ and a distribution $\rho_n$ such that
$$
  \int  KL(P_{\beta_0},P_{\beta}) \rho_n({\rm d}\beta)  \leq\varepsilon_n
  \text{; and } \,
   KL(\rho_n,\pi)  \leq n\varepsilon_n.
$$
 Then, for any $\alpha\in(0,1)$,
$$
\mathbb{E} \left[ \int D_{\alpha}(P_{\beta},P_{\beta_0}) \pi_{n,\alpha}({\rm d}\beta|X ^n) \right]
\leq \frac{1+\alpha}{1-\alpha}\varepsilon_n.
$$
\end{theorem}

\begin{theorem}[Corollary 2.5 in \cite{alquier2020concentration}]
\label{thm_concentration_alquier}
Suppose there is a sequence $\varepsilon_n > 0$ for which a distribution $\rho_n$ exists such that
$$
  \int  KL(P_{\beta_0},P_{\beta}) \rho_n({\rm d}\beta) \leq \varepsilon_n 
  \text{; }
  \int \mathbb{E} \log^2 \left( \frac{p_{\beta}(X_i)}{p_{\beta_0}(X_i)} \right)   \rho_n({\rm d}\beta) \leq \varepsilon_n
 \text{; } \,
   KL(\rho_n,\pi) \leq n \varepsilon_n.
$$
 Then, for any $\alpha\in(0,1)$, 
  \begin{equation*}
  \mathbb{P}\left[
 \int D_{\alpha}(P_{\beta},P_{\beta_0}) \pi_{n,\alpha}({\rm d}\beta|X ^n)  \leq  \frac{2(\alpha+1)}{1-\alpha} \varepsilon_n
 \right]
 \geq 1-\frac{2}{n\varepsilon_n}
 .
 \end{equation*}
\end{theorem}

\begin{lemma}[Lemma 3 in \cite{mai2024concentration}]
	\label{lm_bound_prior_horseshoe}
	Suppose $ \beta_0 \in \mathbb{R}^p $ such that $ \|\beta_0\|_0 = s^* $ and  that $ s^* < n < p $ and $ \|\beta_0\|_\infty \leq C_1  $. Suppose $\beta \sim \pi_{HS} $. 
    Define $\delta=\{s^*\log (p/s^*)/n\}^{1/2} $. Then we have, for some constant $K>0 $, that
$$
\pi_{HS} ( 
\|\beta -\beta_0\|_2<\delta )
	\geq 
	e^{-Ks^*\log (p/s^*)}.
	$$
\end{lemma}

\clearpage
\section{Additional results for simulations}

\begin{figure}[!ht]
    \centering
    \includegraphics[width=12cm]{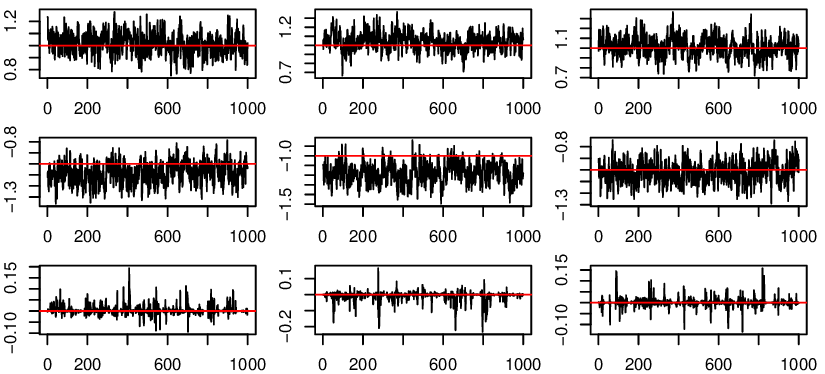}
    \caption{Trace plots from the Gibbs sampler for selected parameter entries.
Top row: three randomly chosen entries with true value 1.
Middle row: three randomly chosen entries with true value $-1$.
Bottom row: three randomly chosen entries with true value 0. Red lines show the true values. }
    \label{fig_tracplot}
\end{figure}

\begin{figure}[!ht]
    \centering
    \includegraphics[width=12cm]{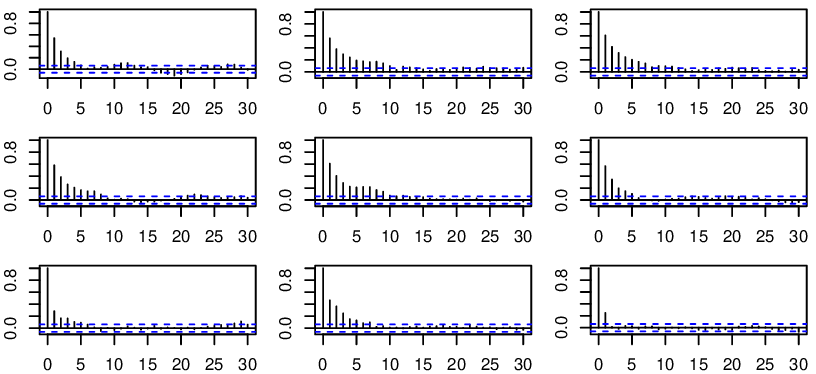}
    \caption{ACF plots from the Gibbs sampler for some random entries as in Figure \ref{fig_tracplot}. Top row (3 plots): 3 random entries with true value 1. Middle row (3 plots): 3 random entries with true value $-1$. Bottom row (3 plots): 3 random entries with true value 0.}
    \label{fig_acf_plot}
\end{figure}

To further demonstrate the behavior of the Gibbs sampler, we present the trace and autocorrelation function (ACF) plots in Figures \ref{fig_tracplot} and \ref{fig_acf_plot}. These results correspond to a simulation scenario with parameters $p = 300$, $n = 200$, $s^* = 10$, and $\rho_X = 0$.

\end{document}